\documentclass[12pt]{article}
\usepackage{amsmath, amssymb}

\newcommand\bth{\ensuremath{\bar{\theta}}}

\newcommand\bi{\ensuremath{\bar{i}}}
\newcommand\bj{\ensuremath{\bar{j}}}
\newcommand\ba{\ensuremath{\bar{a}}}
\newcommand\bb{\ensuremath{\bar{b}}}

\newcommand\bd{\ensuremath{\bar{d}}}

\newcommand\barf{\ensuremath{\bar{f}}}
\newcommand\bk{\ensuremath{\bar{k}}}

\newcommand\bz{\ensuremath{\bar{z}}}

\newcommand\rhF{\ensuremath{\tilde{F}}}
\newcommand\rhH{\ensuremath{\tilde{H}}}

\newcommand\mG{\ensuremath{\mathcal{G}}}

\begin{document}
\vspace*{-.6in} \thispagestyle{empty}
\begin{flushright}
UK-04-22
\end{flushright}
\baselineskip = 18pt

\vspace{1.0in} {\Large
\begin{center}
On Ricci flat supermanifolds
\end{center}} \vspace{.5in}

\begin{center}
Chengang Zhou \\
\emph{Department of Physics and Astronomy\\
University of Kentucky, Lexington, KY  40506, USA\\
czhou@pa.uky.edu}
\end{center}
\vspace{0.4in}

\begin{center}
\textbf{Abstract}
\end{center}
\begin{quotation}
\noindent We study the Ricci flatness condition on generic
supermanifolds. It has been found recently that when the fermionic
complex dimension of the supermanifold is one the vanishing of the
super-Ricci curvature implies the bosonic submanifold has
vanishing scalar curvature. We prove that this phenomena is only
restricted to fermionic complex dimension one. Further we
conjecture that for complex fermionic dimension larger than one
the Calabi-Yau theorem holds for supermanifolds.
\end{quotation}

\newpage

\pagenumbering{arabic}

Calabi-Yau compactification has been one of the most important
cornerstones of superstring phenomenology. It is a supersymmetric
vacuum solution to string theory in the absence of RR and NS-NS
fluxes. Much progress has been made in studying sigma models,
topological models and branes on Calabi-Yau backgrounds.

The extension to Calabi-Yau supermanifold had been attempted
several years ago~\cite{S, Sch}, and received much attention
recently after Witten's proposal that the perturbative amplitudes
of the $\mathcal{N}=4$ super Yang-Mills theory can be recovered
from open string theory on the Calabi-Yau supermanifold
$\mathbb{CP}^{3|4}$~\cite{W}. It belongs to a class of
supermanifolds which can be obtained starting from a certain
bosonic vector bundle over a K\"{a}hler manifold and then
fermionizing the bundle direction. The global holomorphic top form
exists as long as the base manifold and the vector bundle have the
same canonical line bundle. It is reasonable to compactify the
string theory on supermanifolds and look for conformal
backgrounds.

By the famous Calabi-Yau theorem, for given complex structure and
K\"{a}hler class on a K\"{a}hler manifold, there exists a unique
Ricci flat metric if and only if the first Chern class of the
manifold vanishes, or there is a globally defined holomorphic top
form on the manifold. Since the worldsheet sigma model is
conformal invariant only when the target space is Ricci flat, it
follows that the above-mentioned class of supermanifolds are all
valid perturbative string theory backgrounds.

It is then a surprise that R\v{o}cek and Wadhwa~\cite{rw} found a
counterexample to the Calabi-Yau theorem when the supermanifolds
are constructed by fermionizing a line bundle over the base
manifold. They proved that in this case, the super-Ricci flatness
actually requires more than just the vanishing of the first Chern
class: it also requires the bosonic base manifold to have
vanishing scalar curvature. The novelty of the supermanifold
compared to the bosonic manifold, regarding the Ricci flatness, is
that in addition to the vanishing of first Chern class as an
integrability condition, there are local constraints from the
fermionic expansion of the curvature, as we will analyze later.

We will retain the name Calabi-Yau manifolds for K\"{a}hler
manifolds with vanishing first Chern class, or equivalently with a
global holomorphic top form, in the case of supermanifolds. So by
the result of ~\cite{rw}, $\mathbb{CP}^{3|1}$ is a super
Calabi-Yau manifold, as it has a global holomorphic $(3,1)$ form,
but it is not super Ricci-flat as the base manifold
$\mathbb{CP}^3$ has non-vanishing scalar curvature.

A natural question is whether this counterexample to the
Calabi-Yau theorem is merely an exception restricted to fermionic
complex dimension one, or more general valid for higher fermionic
dimension. We will show in this paper that for the Ricci-flat
metric on supermanifold to imply that the bosonic manifold has
vanishing scalar curvature, the condition of the complex fermionic
dimension being one is not only sufficient but also necessary. An
intuitive explanation can be seen in the following simple example.
Consider a function of $z_0$ and $z_i$. The Taylor expansion in
$z_0$ has infinitely many terms
\begin{equation}
F(z_0, z_i) = f(z_i) + f_1(z_i)z_0+ f_1(z_i)z_0^2 + \cdots.
\end{equation}
However, after fermionizing $z_0$, the expansion will be cut off
in the second order due to the Grassmannian nature of the
fermionic coordinates,
\begin{equation}
F(\theta, z_i) = f(z_i) + f_1(z_i)\theta.
\end{equation}
The natural disappearance of the higher order terms, as will be
shown in this paper, directly leads to the result of ~\cite{rw}.

There are additional local constraints for the various Taylor
expansion coefficients of the supermetric from the Ricci-flatness.
Although it is hard to analyze these conditions, we conjecture
that they do not impose additional global topological constraints
on the base manifold, but merely give the relations among the
coefficients of the fermionic expansion of the supermetric.

\section*{Ricci-flat Bosonic K\"{a}hler manifold and superextension}

To understand the peculiarity of the Calabi-Yau supermanifolds,
let us first review the content of the usual Calabi-Yau theorem.
It concerns the existence of the metric of $SU(N)$ holonomy on a
general K\"{a}hler manifold, which says that the necessary and
sufficient condition for its existence is that the manifold has
vanishing first Chern class. There are equivalent conditions which
we will find useful later. For example, a compact K\"{a}hler
manifold has vanishing first Chern class if and only if the
manifold admits a nowhere vanishing holomorphic top form. Also the
metric of $SU(N)$ holonomy and the Ricci-flat metric are the same
thing, see, for example, p. 439 of ~\cite{GSW}. A K\"{a}hler
manifold has the nice property that its Ricci curvature tensor is
simply
related to the K\"{a}hler metric as 
\begin{equation}
R_{i\bar{j}}=-(\ln \mbox{det}(g))_{,i\bar{j}}.
\end{equation}
So locally, $R_{i\bj}=0$ implies
\begin{align}
\ln \mbox{det}(g)=F(z^i) + \bar{F}(\bar{z}^{\bi}),
\end{align}
for an arbitrary holomorphic function $F(z^i)$. By an appropriate
holomorphic change of coordinates, $F(z)$ can be put into any
form, for example, a constant. So then locally the Ricci-flatness
condition implies that the K\"{a}hler potential obeys:
\begin{equation}
\mbox{det}(g)\equiv \mbox{det}(\partial_i\partial_{\bar{j}}K)=1,
\end{equation}
where $K$ is the K\"{a}hler potential. This is a differential
equation about a scalar function $K$, or an algebraic equation
about the metric tensor, whose solution always exists. However
this is just the local condition, and one has to patch all the
local solutions together into a global one. The global
integrability condition for the Ricci-flatness is then the
vanishing of first Chern class.

However, in the case of the supermanifolds, this local equation
implies additional constraints on the structure of the manifold,
besides the global topological condition. To explain this point,
and to compare to the case of the higher fermionic dimension, let
us study more carefully the case of the fermionic complex
dimension one. We will follow the derivation of ~\cite{rw} in this
case.

In the following we will assume the block form of the supermetric
\begin{equation}
\mathcal{G}=\left( \begin{matrix} A & B \cr C & D
\end{matrix} \right),
\end{equation}
where $A$ and $D$ are bosonic, but $C$ and $D$ are fermionic. The
superdeterminant is defined as
\begin{equation}
\mbox{sdet}(\mathcal{G}) \equiv
\frac{\mbox{detA}}{\mbox{det}(D-CA^{-1}B)} =
\frac{\mbox{det}(A-BD^{-1}C)}{\mbox{det}D}.
\end{equation}

Now for supermanifolds with only one complex fermionic dimension,
the fermionic expansion of the K\"{a}hler potential is
\begin{align}
K=G+F\theta\bar{\theta},
\end{align}
and the expansion of the supermetric follows
\begin{align}
 \mathcal{G}=\begin{pmatrix} G_{,i\bar{j}}+ F_{,i\bar{j}} \theta\bar{\theta}
& F_{,i}\theta \cr F_{,\bar{j}}\bar{\theta} & F
\end{pmatrix}.
\end{align}
One should regard $G_{,i\bar{j}}$ as the K\"{a}hler metric on the
bosonic submanifold, which only depends on $z^i$ and
$\bar{z}^{\bi}$. Now a simple calculation yields
\begin{align}
\begin{split}
\mbox{sdet} (\mathcal{G}) &= \frac{1}{F} \mbox{det} (G_{,i\bar{j}}
+ F_{,i\bar{j}} \theta\bar{\theta}
-\frac{F_{,\bar{j}}F_{,i}}{F}\theta\bar{\theta}) \\
&= \frac{\mbox{det}(G_{,i\bar{j}})}{F}
[1+(G_{,i\bar{j}})^{-1}(F_{,i\bar{j}}-
\frac{F_{,\bar{j}}F_{,i}}{F})\theta\bar{\theta}].
\end{split}
\end{align}
Then $\mbox{sdet}(\mathcal{G})=1$ implies
\begin{align}
F &= \mbox{det}(G_{,i\bar{j}}), \\
0 &= (G_{,i\bar{j}})^{-1}(F_{,i\bar{j}}-
\frac{F_{,\bar{j}}F_{,i}}{F}),
\end{align}
and, using the first equation, one immediately realizes that the
second equation means exactly the scalar curvature of the bosonic
manifold vanishes
\begin{align}
\begin{split}
R \equiv (G_{,i\bar{j}})^{-1}R_{i\bar{j}} (\mbox{bosonic}) =
 (G_{,i\bar{j}})^{-1}[\ln\mbox{det}(G_{,i\bar{j}})]_{,i\bar{j}}
 =(G_{,i\bar{j}})^{-1}(\ln F)_{,i\bar{j}} =0.
\end{split}
\end{align}

Because of the Grassmann nature of the fermionic coordinates, both
the K\"{a}hler potential and the super metric has a cutoff in the
fermionic expansion. Were these fermionic coordinates bosonic and
one does the same expansion, there would have been an infinite
number of equations for an infinite number of coefficients of the
expansion. One can then interpretes them as determining the higher
order expansion coefficients in terms of the lower order ones,
which essentially encodes the trivial local equation that we
mentioned above. However, in the fermionic case, the Grassmann
nature imposes an a priori condition that infinitely many
coefficients have to vanish. In turn they imposes consistency
conditions on the lower order expansion coefficients if one still
require it to be the solution. They will appear as additional
constraints on the bosonic submanifold, as we have clearly seen
from the example of the complex fermionic dimension one.

This leads to the conclusion of ~\cite{rw} that the Calabi-Yau
theorem does not hold for the supermanifolds with one fermionic
complex dimension one. In this case, the Ricci-flat supermetric
requires not only the vanishing of the first Chern class, or
equivalent the existence of the global holomorphic top form, but
also requires that the base manifold has vanishing scalar
curvature. For example, $\mathbb{CP}^{1|1}$ has global top
holomorphic form, but does not admit flat Ricci tensor.

\section*{Fermionic complex dimension two and higher}

We will see that the vanishing of the scalar curvature of the
bosonic submanifold from flat super-Ricci tensor is restricted to
complex fermionic dimension one. First let us study the case of
fermionic complex dimension two. Let the expansion of the Kahler
potential be\footnote{Henceforth we will use the notation
$\theta^4$ for $\theta^1\bar{\theta^1}\theta^2\bar{\theta^2}$.}
\begin{equation}
K(z, \bar{z}, \theta, \bar{\theta}) =
G(z,\bar{z})+\sum_{a\bb}F_{a\bb}\theta^a\bar{\theta}^{\bar{b}}+H(z,\bar{z})\theta^4.
\end{equation}
Then the supermetric has the following block form
\begin{equation}
\mathcal{G}=\left( \begin{matrix} A_{i\bar{j}} &
B_{i\bar{\theta}^{\bar{b}}} \cr C_{\theta^a \bar{j}} &
D_{\theta^a\bar{\theta}^{\bar{b}}} \end{matrix} \right),
\end{equation}
where\footnote{The convention for the holomorphic superderivative
is from the left and the anti-holomorphic derivative is from the
right. In particular, there is no minus sign in B.}
\begin{align}
A_{i\bar{j}} &= G_{,i\bj} +\sum_{a\bb}
F_{a\bb,i\bj}(z,\bz)\theta^a\bth^{\bb}
+ H_{,i\bj}(z,\bz)\theta^1\bth^1 \theta^2\bth^2, \\
B_{i\bth^{\bb}} &= \sum_a F_{a\bb,i}(z,\bz)\theta^a + H_{,i}(z,\bz)\hat{\theta^4}^{\bb}, \\
C_{\theta^a \bar{j}} &= \sum_{\bb} F_{a\bb,\bj}(z,\bz)\bth^{\bb} + H_{,\bj}(z,\bz)\hat{\theta^4}^a, \\
\begin{split}
D_{\theta^a\bth^{\bb}} &= F_{a\bb}(z,\bz) + H(z,\bz)
\hat{\theta^4}^{a\bb}, \\
    &=F_{a\bb}(z,\bz) + H(z,\bz) \epsilon_{a\bb
    c\bd}\theta^c\bth^{\bd}.
\end{split}
\end{align}
Here the notation $\hat{\theta^4}^{\bb}$ means removing the
$\bth^{\bb}$ from $\theta^4\equiv \theta^1\bth^1 \theta^2\bth^2$,
and likewise for $\hat{\theta^4}^a$ and $\hat{\theta^4}^{a\bb}$.
In the following, we let $g_{i\bj} \equiv
\partial_i\bar{\partial}_{\bj}G$ denote the K\"{a}hler metric on the purely bosonic
submanifold of the supermanifold.

First we compute the fermionic expansion of the determinant of
$A_{i\bar{j}}$. Utilizing the general expansion formula
\begin{align}
\begin{split}
\mbox{det}(I+\Gamma) & = \exp [\mbox{tr} \ln (I+\Gamma)]\\
& = 1+ \sum_a\Gamma_{aa} + \frac{1}{2} \sum_{a,b}
(\Gamma_{aa}\Gamma_{bb} - \Gamma_{ab}\Gamma_{ba}) +\cdots \\
\end{split}
\end{align}
it is straightforward to find
\begin{align}
\mbox{det} A = \mbox{det}(g_{i\bj})[1+\mbox{tr}_{i\bj}
\tilde{F}_{a\bb}\theta^a\bth^{\bb}
    + (\mbox{tr}_{i\bj} \tilde{H} +\mbox{det}_{a\bb} \mbox{tr}_{i\bj}
    \tilde{F} -\mbox{tr}_{i\bj} \mbox{det}_{a\bb}\tilde{F}) \theta^4].
\end{align}

The notations used here needs some explanations. The double
derivatives of $F_{a\bb}$ can be regarded as a tensor of matrices,
with $a\bb$ and $i\bj$ as two pairs of matrix indices. $g_{i\bj}
\equiv \partial_i\bar{\partial}_{\bj}G$ is the metric on the
bosonic submanifold, and we use its inverse to raise the indices
on $i\bj$. Then we define the following simplified notations
\begin{align}
(\tilde{F}_{a\bb})^{\bi}_{\bj}& \equiv g^{\bi k}F_{a\bb,k\bj}(z,\bz), \\
\tilde{H}^{\bi}_{\bj} & \equiv g^{\bi k} H_{,k\bj}(z,\bz),
\end{align}
where $\tilde{H}^{\bi}_{\bj}$ is a matrix in $\bi\bj$ indices, and
$(\tilde{F}_{a\bb})^{\bi}_{\bj}$ is a double matrix. There is
difference in the order of taking matrix operations in the two
pairs of indices, and we have used the following notations
\begin{align}
 \mbox{tr}_{i\bj} \mbox{det}_{a\bb}\tilde{F} &\equiv
\frac{1}{2}\sum_{a\bb c\bd} \epsilon_{a\bb c\bd}
\rhF_{a\bb,\bj}^{\bi} \rhF_{c\bd,\bj}^{\bi}, \\
\mbox{det}_{a\bb} \mbox{tr}_{i\bj} \tilde{F} &\equiv
\frac{1}{2}\sum_{a\bb c\bd} \epsilon_{a\bb c\bd}
(\sum_{\bi}\rhF_{a\bb,\bi}^{\bi}) (\sum_{\bj}
\rhF_{c\bd,\bj}^{\bj}).
\end{align}
For example, $\mbox{tr}_{i\bj} \mbox{det}_{a\bb}\tilde{F}$ means
that we first regard $\tilde{H}^{\bi}_{\bj}$ as a matrix labelled
by $a\bb$ with each matrix element a matrix labelled by
$(\bi,\bj)$ itself, and take the determinant in $a\bb$ using the
matrix multiplication in $(\bi,\bj)$ when we multiplying the
matrix elements. The result determinant is a matrix itself in
$(\bi,\bj)$ and one can take the trace over it. Similar
explanation holds for $\mbox{det}_{a\bb} \mbox{tr}_{i\bj}
\tilde{F}$.

To compute $\mbox{det}^{-1}(D-CA^{-1}B)$, one first has
\begin{align}
A^{-1} = (1-\rhF_{a\bb}\theta^a\bth^{\bb}- \rhH\theta^4
+2(\mbox{det}_{a\bb}\rhF)^{\bi}_{\bj} \theta^4 )\cdot g^{-1},
\end{align}
then the expansion $CA^{-1}B$ follows
\begin{align}
\begin{split}
(CA^{-1}B)_{a\bb} &\equiv C_{\theta^a \bi} (A^{-1})^{\bi j}
                     B_{j\bth^{\bb}} \\
        & = -\sum_{c\bd} F_{a\bd,\bi}g^{\bi j}F_{c\bb,j}
        \theta^c\bth^{\bd} - (F_{a\bd,\bi}g^{\bi j}H_{,j}
        + H_{,\bi}g^{\bi j}F_{a\bb,j}- \epsilon^{c\bd e\barf}F_{a\bd,\bi}
        (\rhF_{e\barf})^{\bi}_{\bk} g^{\bk j} F_{c\bb,j}  )\theta^4
\end{split}
\end{align}

The inverse determinant, using the following expansion formula,
\begin{equation}
\begin{split}
\mbox{det}^{-1}(I-\Gamma) & = \exp [-\mbox{tr} \ln (I-\Gamma)]\\
& = 1+ \sum_a\Gamma_{aa} + \frac{1}{2} \sum_{a,b}
(\Gamma_{aa}\Gamma_{bb} + \Gamma_{ab}\Gamma_{ba}) +\cdots \\
\end{split}
\end{equation}
becomes
\begin{align}
\begin{split}
\mbox{det}^{-1}(D-CA^{-1}B) & =  (\mbox{det} F_{a\bb})^{-1} \{1-
       F^{\ba b}(H\epsilon_{b\ba c\bd} + F_{b\bd,\bi}
       g^{\bi j} F_{c\ba,j})\theta^c \bth^{\bd} \\
    & \quad + [-g^{\bi j} \mbox{tr}(F^{-1}\partial_{\bi} F) \partial_{j}H -g^{\bi j}
     \mbox{tr}(F^{-1}\partial_{j}F) \partial_{\bi}H \\
    & \quad +\epsilon^{a\bb c\bd}(g^{\bj k}\partial_k FF^{-1} \partial_{\bi} F)_{a\bb}
     (g^{\bi l}\partial_l\partial_{\bj}F)_{c\bd} \\
    & \quad +H g^{\bi j}\mbox{tr}^{\ba}_{\bb}(F^{-1}\partial_{\bi}FF^{-1}\partial_jF)
    + H g^{\bi
       j}\mbox{tr}(F^{-1}\partial_{\bi}F)\mbox{tr}(F^{-1}\partial_{j}F)
       \\
    & \quad + \mbox{det}_{a\bb} \mbox{tr}_{i\bj}
    (\partial_jFF^{-1}\partial_{\bi}
F) - \mbox{tr}_{i\bj} \mbox{det}_{a\bb} (g^{\bi k}\partial_k
FF^{-1}\partial_{\bj}F)] \theta^4\}.
\end{split}
\end{align}

Finally, putting everything together, we obtain the
superdeterminant of the super metric $\mG$
\begin{align}
\begin{split}
\mbox{sdet}\mG & = \frac{\mbox{det}g}{\mbox{det}F} \{1+ (g^{\bj
        i}F_{a\bb, i\bj} - H F^{\bd c}\epsilon_{c\bd a\bb} - F^{\bd c}
        F_{c\bb, \bj} g^{\bj i} F_{a\bd,i}) \theta^a\bth^{\bb} \\
    & \quad + [\mbox{tr}\tilde{H} + \mbox{det}_{a\bb} (\mbox{tr}_{i\bj}
     \tilde{F})-\mbox{tr}_{i\bj}(\mbox{det}_{a\bb}\tilde{F}) \\
    & \quad + g^{\bi j}\mbox{tr}(F^{-1}\partial_{\bi}\partial_j
    F)H -\epsilon^{a\bb c\bd} (\mbox{tr}_{i\bj}\tilde{F}_{a\bb})
    (g^{\bi j} \partial_jFF^{-1}\partial_{\bi}F)_{c\bd} \\
    & \quad -g^{\bi j} \mbox{tr}(F^{-1}\partial_{\bi} F) \partial_{j}H -g^{\bi j}
     \mbox{tr}(F^{-1}\partial_{j}F) \partial_{\bi}H \\
    & \quad +\epsilon^{a\bb c\bd}(g^{\bj k}\partial_k FF^{-1} \partial_{\bi} F)_{a\bb}
     (g^{\bi l}\partial_l\partial_{\bj}F)_{c\bd} \\
    & \quad +H g^{\bi j}\mbox{tr}^{\ba}_{\bb}(F^{-1}\partial_{\bi}FF^{-1}\partial_jF)
    + H g^{\bi j}\mbox{tr}(F^{-1}\partial_{\bi}F)\mbox{tr}(F^{-1}\partial_{j}F) \\
    & \quad + \mbox{det}_{a\bb} \mbox{tr}_{i\bj}
    (\partial_jFF^{-1}\partial_{\bi}
F) - \mbox{tr}_{i\bj} \mbox{det}_{a\bb} (g^{\bi k}\partial_k
FF^{-1}\partial_{\bj}F)] \theta^4 \}.
\end{split}
\end{align}

The condition for Ricci-flatness, $\mbox{sdet}\mG =1$, implies the
following set of equations,
\begin{align}
\mbox{det}g & =\mbox{det}F,  \label{EQA} \\
0 &= -HF^{\bd c}\epsilon_{c\bd a\bb} + g^{\bj i}
(F_{a\bb,i\bj}-F_{a\bd,i}
F^{\bd c} F_{c\bb,\bj}) ,  \label{EQB} \\
\begin{split}
0 & = \mbox{tr}\tilde{H} + \mbox{det}_{a\bb} (\mbox{tr}_{i\bj}
     \tilde{F})-\mbox{tr}_{i\bj}(\mbox{det}_{a\bb}\tilde{F}) \\
    & \quad + Hg^{\bi j}\mbox{tr}(F^{-1}\partial_{\bi}\partial_j F)
    -\epsilon^{a\bb c\bd} \mbox{tr}_{i\bj}\tilde{F}_{a\bb}
    (g^{\bi j} \partial_jFF^{-1}\partial_{\bi}F)_{c\bd} \\
    & \quad -g^{\bi j} \mbox{tr}(F^{-1}\partial_{\bi} F) \partial_{j}H -g^{\bi j}
     \mbox{tr}(F^{-1}\partial_{j}F) \partial_{\bi}H \\
    & \quad +\epsilon^{a\bb c\bd}(g^{\bj k}\partial_k FF^{-1} \partial_{\bi} F)_{a\bb}
     (g^{\bi l}\partial_l\partial_{\bj}F)_{c\bd} \\
    & \quad +H g^{\bi j}\mbox{tr}^{\ba}_{\bb}(F^{-1}\partial_{\bi}FF^{-1}\partial_jF)
    + H g^{\bi j}\mbox{tr}(F^{-1}\partial_{\bi}F)\mbox{tr}(F^{-1}\partial_{j}F) \\
    & \quad + \mbox{det}_{a\bb} \mbox{tr}_{i\bj}
    (\partial_jFF^{-1}\partial_{\bi}
F) - \mbox{tr}_{i\bj} \mbox{det}_{a\bb} (g^{\bi k}\partial_k
FF^{-1}\partial_{\bj}F).
\end{split}
\label{EQC}
\end{align}

To figure out the geometrical meaning of these equations, it is
best to regard these as matrix equations. Multiplying the second
equation by $g^{\bb a}$ and taking trace,
\begin{equation}
g^{\bj i}[\mbox{tr} F^{-1}\partial_i\partial_{\bj} F
-\mbox{tr}(F^{-1}\partial_iF F^{-1} \partial_{bj} F)] - 2H
\mbox{det} F^{-1}=0.
\end{equation}
It is easy to verify the following equality for matrix $F$
\begin{equation}
\mbox{tr}\partial_i\partial_{\bj}(\ln F) =
\mbox{tr}(F^{-1}\partial_i\partial_{\bj} F - F^{-1}\partial_i F
F^{-1}
\partial_{\bj} F),
\end{equation}
from which one has
\begin{equation}
-2H\mbox{det}^{-1}F = -g^{\bj
i}\partial_i\partial_{\bj}(\mbox{tr}\ln F).
\end{equation}
Because $\mbox{det}F =\mbox{det}g$, the righthand side of the
equation is exactly the Ricci scalar curvature $R\equiv -g^{\bj i}
\partial_i \partial_{\bj}\mbox{tr}\ln g$ of the bosonic base manifold. So one
finally finds out the scalar curvature for the bosonic submanifold
\begin{align}
R=-2H\mbox{det}^{-1}g. \label{EQBA}
\end{align}
$g$ is the metric of the bosonic manifold which is certainly
non-degenerate, $\mbox{det}g\neq 0$. So as long as $H\neq 0$, the
bosonic base manifold need not have vanishing scalar curvature.

$H$ is the coefficient of the quartic term in the fermionic
expansion, which naturally disappears if there is only one complex
fermionic dimension as studied in~\cite{rw}. It is exactly the
finite cutoff determined by the dimension of the fermionic degrees
of freedom in the fermionic expansion causes the anomaly in the
dimension one.

When the complex fermionic dimension is larger than two, one has
similar result concerning the scalar curvature for the bosonic
submanifold. The K\"{a}hler potential has expansion in fermionic
coordinates
\begin{align}
K(z, \bar{z}, \theta, \bar{\theta}) =
G(z,\bar{z})+\sum_{a\bb}F_{a\bb}\theta^a\bar{\theta}^{\bar{b}}+
\sum_{a\bb c\bd}H_{a\bb c\bd}(z,\bar{z}) \theta^a\bth^{\bb}
\theta^c\bth^{\bd}+\cdots.
\end{align}
where the scalar $H$ in the complex fermionic dimension two has
been replaced by a totally anti-symmetric tensor $H_{a\bb c\bd}$.
Up to $\theta\bth$ order in the superdeterminant expansion there
are similar equations
\begin{align}
\mbox{det}g & =\mbox{det}F, \\
F^{\bd c}H_{c\bd a\bb} &= g^{\bj i} (F_{a\bb,i\bj}-F_{a\bd,i}
F^{\bd c} F_{c\bb,\bj}).
\end{align}
The scalar curvature of the bosonic manifold is then
\begin{align}
R=- F^{\bd c}F^{\bb a}H_{c\bd a\bb}.
\end{align}
It is obvious that the scalar curvature can generally take any
value. This proves our claim.

However, one can not conclude that Calabi-Yau theorem holds for
general supermanifold when the fermionic complex dimension is two
or higher yet. The reason is that there are equations from the
higher order expansions of the superdeterminant, and one may
wonder if they impose topological or geometrical conditions on the
bosonic submanifold. We will study the case for fermionic
dimension two and find some hints.

Use the first two equations to simplify the last equation, and
after a bit of algebra, we find
\begin{align}
\mbox{tr}_{i\bj} \mbox{det}_{a\bb}(Fg^{\bi
k}\partial_k\partial_{\bj}\ln
    F) = H (g^{\bi j} \partial_{\bi} \ln R\partial_j
    \ln R+ H g^{\bi j} \partial_{\bi} \partial_j \ln H
    +\frac{R}{2}).
\end{align}
Notice that using the previous two equations and the expression
for $H$, the righthand side of the equation is determined by the
metric of the bosonic submanifold alone. Now after solving for
$H$, one has five equations on $F_{a\bb}$ in terms of $g_{i\bj}$,
one from ~\eqref{EQA}, three from ~\eqref{EQB} after using
~\eqref{EQBA} and one from ~\eqref{EQC}. Naively this should lead
to one consistency condition for $g_{i\bj}$, however notice that
some of these are nonlinear second order differential equations,
and such simple counting may be wrong.

One can look at some simple examples. For example, the
supermanifold $\mathbb{CP}^{1|2}$ is super-Ricci flat. Actually
the super-Ricci flat metric is the Fubini-Study metric, which is
from the K\"{a}hler potential
\begin{align}
K(z,\bar{z}, \theta^a,\bth^{\bb}) = \ln(z\bar{z}+
\theta^1\bth^{\bar{1}}+\theta^2\bth^{\bar{2}}).
\end{align}
as one can easily verify.

If one regards $F_{a\bb}$ and their first order derivatives as
independent functions, these five equations become nonlinear first
order equations and there will be more independent functions than
constraints. Then there is no consistency condition and so no
constraints on the bosonic submanifold, and these equations merely
states that $F_{a\bb}$ is correlated to $g_{i\bj}$ for super Ricci
flat metric. Although we have no solid proof, we believe that the
Calabi-Yau theorem does hold for fermionic complex dimension two
or higher. Certainly it will be interesting to either prove or
disprove this conjecture in the future.

\section*{Acknowledgement}

The author would like to thank Jeremy Michelson and Xinkai Wu for
helpful discussions. This work is supported by NSF grant
PHY-0244811 and DOE grant DE-FG01-00ER40899.

\end{document}